\documentclass[11pt,aps,amsmath,amssymb,preprint,nofootinbib,showkeys]{revtex4}
\usepackage[active]{srcltx}
\usepackage[utf8]{inputenc}
\usepackage{latexsym}
\usepackage{amsmath}
\usepackage{graphicx}
\usepackage{slashed}
\usepackage{xcolor}

\begin{document}

\title{Chiral Edge States in 2+1 Dimensional Topological Phases}

\author{Carlos A. Hernaski}
\email{carlos.hernaski@gmail.com}
\affiliation{Departamento de F\'isica, Universidade Estadual de Londrina, 
Caixa Postal 10011, 86057-970, Londrina, PR, Brasil}

\author{Pedro R. S. Gomes}
\email{pedrogomes@uel.br}
\affiliation{Departamento de F\'isica, Universidade Estadual de Londrina, 
Caixa Postal 10011, 86057-970, Londrina, PR, Brasil}

\begin{abstract}

Chiral edge states of 2+1 dimensional Abelian and non-Abelian topological phases can be represented by chiral conformal field theories with integer and non-integer values of central charge, respectively. In this work we describe certain edge states in terms of constrained fermionic fields that realize chiral coset
CFT structures.
This construction arises naturally in the so-called quantum wires approach for topological phases and allows 
for representing fractionalized edge states directly in terms of fermionic degrees of freedom. At the same time, the constrained fermions description introduces some subtleties concerning gauge anomalies since it involves the coupling of chiral fermions to gauge fields. We describe in this article how to handle these issues. 
\end{abstract}

\keywords{Chiral edge states; Topological phases; Quantum wires; Constrained fermions; Chiral CFT.}

\maketitle


\section{Introduction}

\subsection{Motivations}

A remarkable feature of 2+1 dimensional gapped topological phases of matter is the existence of gapless degrees of freedom propagating along their boundaries, which under broad conditions can be described by 1+1 dimensional conformal field theories (CFT) \cite{Wen,Read}. Usually these phases can be classified into Abelian or non-Abelian according to the anyonic statistics of the quasi-particle excitations \cite{Nayak}. From the edge CFT point of view the Abelian phases are characterized by an integer value of central charge, whereas non-Abelian phases possess a fractional contribution, $c=c_{\text{integer}}+c_{\text{noninteger}}$, with $c_{\text{noninteger}}<1$ \cite{Huang}. This work is dedicated to the study of field theories suitable for the description of topological phases exhibiting chiral edge states, i.e., one-way propagating conformal modes at the boundaries.

We should point out about the physical interpretation of 1+1 dimensional models with imbalanced chiral modes. In condensed matter systems, the discrete translational invariance, due to the presence of the lattice, forces the chiral modes to be paired up as a consequence of the Nielsen-Ninomyia theorem \cite{nielsen}. This would jeopardize the interpretation of the chiral field theory as a low-energy effective theory for condensed matter systems, which is precisely the case we are interested in this work. One way out is to consider the 1+1 dimensional theory as the boundary of a 2+1 dimensional one, avoiding the consequences of the Nielsen-Ninomyia theorem. The field theory version of this discussion is related to the Adler-Bell-Jackiw anomaly \cite{Adler,Bell}. It is known that in general a chiral theory cannot be quantized in a gauge invariant way in even dimensions due to the gauge anomaly. The solution to this problem is analogous to the previous one: we need to couple the system to another anomalous system in order to have a cancellation of the gauge anomaly. The Chern-Simons (CS) action in presence of a boundary is the suitable bulk model to adjust the charge conservation. This mechanism is called anomaly-inflow and plays an important role in the CFT-CS bulk-edge correspondence \cite{Wen,Witten5}.

\begin{figure}[!h]
\centering
\includegraphics[scale=0.7]{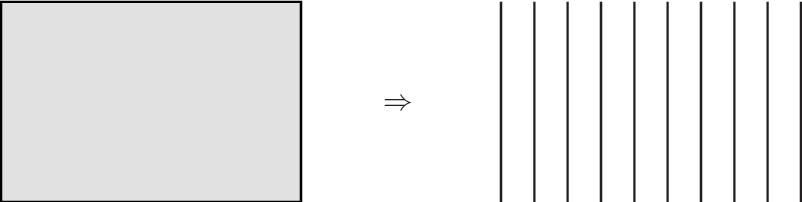}
\caption{Deconstruction of a two dimensional spatial surface in terms of a set of one dimensional systems. Interactions between the wires allow for electrons to jump from one wire to another. This mechanism is expected to capture some essential ingredients of the low-energy physics of the two dimensional phase.}
\label{WiresSet}
\end{figure}

Edge states of topological phases become rather transparent in the quantum wires formalism. It was pioneered by Kane and collaborators in the study of Abelian fractional quantum Hall phases in Ref. \cite{Kane} and it was extended to the non-Abelian case in Ref. \cite{Teo}. Thenceforth, several works have been reported describing successfully certain topological phases in 2+1 dimensions, including those of the tenfold way \cite{Neupert}. A recent comprehensive  review can be found in \cite{Meng1}. The quantum wires system can be interpreted as a dimensional deconstruction of a two dimensional spatial surface by discretizing one of the directions, as depicted in Fig. \ref{WiresSet}. We initially consider a set of fermionic modes propagating along an array of non-interacting wires. The bulk properties of the topological phase are attained by introducing suitable interactions among the wires in such a way to provide a full gap to the bulk modes while leaving gapless some of the edge modes. From a technical point of view, this approach turns the problem into a 1+1 dimensional one, where we have all the non-perturbative machinery of two dimensional quantum field theory as conformal invariance techniques and bosonization.

We proceed by discussing some aspects of CFT that are relevant for our purposes. The Virasoro algebra of a CFT in 1+1 D is suitably constructed from a Kac-Moody algebra of currents in the so-called Sugawara construction (see, for example, Ref. \cite{DiFrancesco}). This approach provides the central charge corresponding to semi-simple affine Lie algebras, which includes the unitary models with $c>1$. The generalization of this construction to other  algebraic structures, as the coset of semi-simple algebras, can be obtained in the Goddard-Kent-Olive (GKO) construction, which can be used to describe CFT with $c<1$ \cite{GKO}.  In this case, one starts with a group $G$ and a proper subgroup $H$. Through the Sugawara construction one defines the energy momentum tensor $T_{G/H}\equiv T_{G}-T_H$, which realizes a Virasoro algebra for the coset structure $G/H$, since there is no singular terms in the operator product expansion between $T_{G/H}$ and $T_H$. The central charge is then given by $c_{G/H}=c_G-c_H$. Within this construction one can obtain a CFT with $c<1$.

Concerning field theory realizations, the Sugawara construction can be faithfully represented in terms of a theory of free fermions belonging to the fundamental representation of a semi-simple group. Non-Abelian bosonization provides a bosonic version of this formalism in terms of the Wess-Zumino-Witten model (WZW) \cite{witten}. In turn, the field theoretical realization of the GKO construction can be implemented in terms of constrained fermions or, in the bosonic version, in terms of gauged WZW models \cite{Polyakov1,Polyakov2,witten2,karabali}. In the constrained fermions approach one starts with a set of free fermions belonging to the fundamental representation of a simple group $G$ and impose that the physical Hilbert space is annihilated by a set of currents belonging to a subgroup $H$. The central charge of the resulting CFT corresponds to that of the coset $G/H$ in the GKO construction. From the WZW model perspective, the constraint can be successfully implemented by gauging the $H$ subgroup of the initial symmetry group $G$.

As firstly noticed in \cite{Hernaski}, the formalism of constrained fermions is closely connected to the quantum wires description of topological phases.  Specifically, the strong coupling limit (low-energy) of the quantum wires is equivalent to impose certain constraints on the initially free fermions, decreasing the conformal content of the theory or, equivalently, giving a gap for some of the modes.  In addition, as we shall discuss, the description in terms of constrained fermions turns out to be an important link to relate the quantum wires to the low-energy effective bulk theory given in terms of Chern-Simons theories.   
In \cite{Hernaski}, we dealt with the case where both right and left chiralities were treated on an equal foot by always keeping the two edges of the system simultaneously in order to form the two-chirality pair of right and left Weyl fermions. In this work we further generalize the construction to the case of chiral constrained fermions, where we can handle with each boundary theory independently. This introduces some subtleties since it involves a coupling of chiral fermions to gauge fields, which in general brings up issues with gauge anomalies.

\subsection{Statement of the Problem and Main Results}

Within the formalism of constrained fermions, the authors in \cite{Cabra} succeeded in obtaining non-chiral CFT, i.e., a CFT whose central charge of left and right sectors are the same, $c_L=c_R$, realizing the general coset structure of affine Lie algebras $G/H$ of the GKO construction. In Ref. \cite{fradkin}, a similar construction was employed, but considering only left fermions from the beginning. As we will review, the constraints can be implemented in the fermionic system by coupling the fermions to auxiliary gauge fields. This is well known to lead to a gauge anomalous theory. Since we cannot restore the gauge invariance choosing a convenient regularization procedure, there is a class of nonequivalent models that we can obtain by adjusting a Jackiw-Rajaraman-type parameter in the quantum theory \cite{Jackiw}. In the work of Ref. \cite{fradkin} the theory is regularized by adding a decoupled set of right fermions and understanding the problem as a gauged fixed model with both chiralities coupled to a vector gauge field. Then, non-Abelian bosonization can be used, but this procedure does not lead to a coset structure for the resulting CFT.

The bosonic counterpart of this issue arises naturally in the attempt to produce general coset structures $G/H_L\times G/H_R$ out of the WZW model by gauging independent subgroups $H_L$ and $H_R$ of $G$. It is known that due to the non-Abelian anomaly, the most general subgroup of $G$ that can be gauged in the WZW model must satisfy $H_L=H_R$ \cite{witten2}. The gauge symmetry is then called vector gauge symmetry, and we cannot produce chiral CFT using this procedure.
It is worth mentioning that only in two dimensions one can have a more general solution to the non-Abelian anomaly cancellation by letting the gauge fields $A_L$ and $A_R$ transform with gauge independent parameters $\theta_L(x^-)$ and $\theta_R(x^+)$ depending only on half of the light cone coordinates \cite{Tye}. This seems to be appropriate to describe chiral coset CFT. However, as noticed in \cite{Tseytlin}, this symmetry is not a true gauge symmetry, since it cannot be used to suppress dynamical degrees of freedom. It is shown in \cite{Tseytlin} that this chiral gauging of the WZW model results in a chiral CFT, but it does not correspond to the coset structure $G/H_L\times G/H_R$. We will come back to this point later.

In this work we shall discuss that in order to obtain a coset structure in the general chiral case from a system of quantum wires, one should use the prescription of identifying the ill-defined chiral determinants $\det(D_{\pm})$ with the so-called chiral WZW models \cite{Sonnenschein} and not with the standard WZW action. This identification is equivalent to the problem of chiral bosonization \cite{Sonnenschein}. With this procedure, we will be able to describe conformal field theories that realize the coset structure $G/H_L\times G/H_R$, i.e., with central charges for the right and left sectors given by $c_R=c_G-c_{H_R}$ and $c_L=c_G-c_{H_L}$, respectively. As we have mentioned, the system of constrained fermions arises naturally in the quantum wires approach for topological phases and it enables to represent fractionalized edge states directly in terms of fermionic degrees of freedom. By using this approach we will illustrate how to generate classes of chiral CFT including the series of minimal and superconformal models.

This work is organized as it follows. In Sec. \ref{S1} we discuss the connection between the interacting quantum wires and the constrained fermions. 
In Sec. \ref{S2} we discuss how to obtain the desired chiral coset structures via chiral bosonization with an appropriate regularization for chiral fermionic 
determinants. The connection with bulk Chern-Simons theory is discussed in Sec. \ref{BulkEdge}. In Sec. \ref{S3} we apply our construction to the description of edge states of some classes of topological phases. We then conclude the work with a summary and some remarks in Sec. \ref{fr}. There is also an appendix where we discuss some details of the interactions involved in the quantum wires construction. 

\section{From Quantum wires to Constrained fermions\label{S1}}

In this section we will discuss how the formalism of constrained fermions naturally emerges in the description of the strong coupling limit of the system of quantum wires.

To understand the underlying ideas, we consider a system of massless fermions propagating along an array of initially decoupled one dimensional wires:
\begin{equation}
 \mathcal{L}_0=\sum_{I=1}^N \mathrm{i} \left(\psi_{R,I}^{\ast}\partial_{+}\psi_{R,I}+\psi_{L,I}^{\ast}\partial_{-}\psi_{L,I}\right),\label{free}
\end{equation}
where $\partial_{\pm}\equiv \partial_t\pm\partial_x$. The discrete label $I=1,...,N$ specifies in which wire the fermions are propagating, but also can be seen as an internal index of a symmetry group $G$, for which the set of free fermions in 1+1 dimensions forms a representation. Initially, the decoupled set of wires describes a CFT with central charge proportional to the number of wires, which counts the number of gapless degrees of freedom. 

As mentioned previously, the strategy to produce a 2+1 dimensional topological phase is to introduce interactions between neighboring wires\footnote{The interactions can involve two or more wires.}. Let us consider the interactions generically as $\sum_{a}\lambda_a\mathcal{L}_{int}^a$, with $\lambda_a$ being the coupling constants. To obtain a stable topological phase the interactions should be such that they provide a full gap to the bulk modes while leaving gapless some of the modes of the borders. This means that the corresponding coupling constants must be relevant, i.e., they need to flow to a strongly coupled regime at low energies. Several constructions where these conditions are met can be found in Refs. \cite{Teo,Fuji} for the case of quantum Hall phases and in Refs. \cite{Sela,Meng,Huang} for spin liquid phases.

We will focus on topological phases whose edge states are represented by conformal field theories involving coset structures of semi-simple algebras, which can produce theories with central charge smaller than one. Thus, it is natural to try to associate gapped and gapless sectors to symmetry groups. In this case, the strategy is to introduce interactions between neighboring wires involving massless modes pertaining to representations of a subgroup $H$ of the initial symmetry group $G$. Thus, if the interactions fulfill the above requirements we expect that the remaining massless modes located at the boundaries realize a CFT with central charge corresponding to the coset structure $G/H$. 

Under the above conditions the ground state properties of the topological phase can be accessed by projecting  out the gapped modes of the spectrum of the theory. This is equivalent to impose that the physical spectrum is a singlet with respect to the subgroup $H$, i.e., given the set of $H$-currents, $J^A$, we impose
\begin{equation}
J^A\left|\text{phys}\right.\rangle=0,\label{e2}
\end{equation}
with $A=1,\ldots,\text{dim}(\hat{H})$, where $\hat{H}$ is the algebra of $H$. 
In a path integral formulation these conditions can be implemented by including appropriate $\delta$-functionals in the partition function:
\begin{equation}
Z=\int{\mathcal{D}\bar{\psi}\mathcal{D}\psi\prod_{A}\delta(J^{A}(x))e^{\mathrm{i}\int{d^2x\sum\bar{\psi}^{I}\mathrm{i}\slashed\partial\psi^{I}}}}.\label{e3}
\end{equation}
The 2+1 dimensional topological phase is realized with the bulk completely gapped and the edges supporting central charges corresponding to coset theories. The edge quantum theory can then be identified with a constrained fermionic model.
The identification of the strong coupling limit of the interacting quantum wires system with a constrained fermion theory is an important link for the connection of the quantum wires approach with the more abstract Chern-Simons description of the topological phases, since both can be matched to the same edge theory through the bulk-edge correspondence \cite{Hernaski,Rossini}.

To identify the edge CFT described by (\ref{e3}) we rewrite the constraints using Lagrange multiplier fields $A^A$ taking values in the Lie-algebra of $H$:
\begin{equation}
Z=\int{\mathcal{D}\bar{\psi}\mathcal{D}\psi\mathcal{D}A\exp{\mathrm{i}\int{d^2x\left(\bar{\psi}\left(\mathrm{i}\slashed\partial+\slashed A\right)\psi\right)}}}.\label{7}
\end{equation}
This theory describes fermions coupled to dynamical gauge fields.
Through standard procedures one can generally rewrite such a theory in terms of free fermions, ghost systems and WZW models, whose central charges can be promptly calculated. By following this strategy, the authors in \cite{Cabra} considered the problem of producing coset CFTs in 1+1 dimensions from systems of constrained fermions. They were able to obtain CFT models compatible with the coset structure of simple algebras $G/H$ of the GKO construction. Since they treated the left and right sectors equally, the 1+1 dimensional CFT obtained describes equal numbers of left and right modes, i.e., $c_L=c_R$. Seen as a boundary of a 2+1 dimensional topological phase, this situation corresponds to a non-chiral or time-reversal invariant phase.

The description of chiral CFT, i.e., $c_L\neq c_R$, which is essential to treat chiral topological phases, is however more subtle. Indeed, it is known that a chiral theory coupled to a gauge field cannot be quantized in a gauge invariant way in even dimensions due to the Adler-Bell-Jackiw. Nevertheless, this fact does not prevent the existence and the successful description of chiral gapped topological phases in 2+1 dimensions. In the CS description, for example, the anomaly in the edge theory is compensated by the also anomalous CS theory in the presence of a boundary. The chiral edge theory can then be described by a gauged chiral WZW model realizing the GKO construction. However, as discussed above, starting from the quantum wires approach the edge CFT is naturally described in terms of constrained fermions and the GKO construction is not obtained straightforwardly, as can be noticed from the results obtained in Ref. \cite{fradkin}. We shall follow in this work an alternative path and give a different prescription for the chiral determinants $\det(D_{\pm})$ that is consistent with the non-Abelian bosonization of both Weyl and Dirac fermions.

\section{Chiral coset CFT from constrained fermions}\label{S2}

To understand the subtleties involved in the description of chiral CFTs, we generalize our discussion of the previous section and consider the problem of constraining independent subgroups $H_L$ and $H_R$ of the symmetry group $U(N)_L\times U(N)_R$ of the Lagrangian (\ref{free}). Denoting by $J^A_L$ and $J^A_R$ the conserved $H_L$ and $H_R$-currents, respectively, and by $A_L$ and $A_R$ the auxiliary Lie-algebra-valued gauge fields, the constrained partition function of the boundary theory can then be written as
\begin{eqnarray}
Z&=&\int{\mathcal{D}\psi_L\mathcal{D}\psi_R\mathcal{D}A^L\mathcal{D}A^R}\nonumber\\
&\times&\exp{\mathrm{i}\int{d^2x\left(\psi^{i\alpha\ast}_L\left(\mathrm{i}\delta^{\alpha\beta}\partial_-+\left(A^L_-\right)^{\alpha\beta}\right)\psi^{i\beta}_L+\psi^{i^{\prime}\alpha^{\prime}\ast}_R\left(\mathrm{i}\delta^{\alpha^{\prime}\beta^{\prime}}\partial_++\left(A^R_+\right)^{\alpha^{\prime}\beta^{\prime}}\right)\psi^{i^{\prime}\beta^{\prime}}_R\right)}},\label{7.1}
\end{eqnarray}
where we have broken the $N$ left (right) fermions in the fundamental representation of $U(N)_L$ ($U(N)_R$) into $N^L_f$ ($N^R_f$) fundamental representations of the subgroup $H_{L}$ ($H_R$). The new fermion indexes have the following structure: $i=1,\ldots,N^L_f$, $i^{\prime}=1,\ldots,N^R_f$, $\alpha,\beta=1,\ldots,N^L_c$, and $\alpha^{\prime},\beta^{\prime}=1,\ldots,N^R_c$, with $N^L_c$ ($N^R_c$) being the number of colors transforming under the subgroups $H_{L}$ ($H_R$). Then, $N^R_fN^R_c=N^L_fN^L_c=N$.

To state the problem we temporarily let the gauge fields be external and 
consider the effective action $W[A^L,A^R]$, which is defined by
\begin{equation}
e^{\mathrm{i}W[A^L,A^R]}=\int{\mathcal{D}\psi_L\mathcal{D}\psi_R\exp{\mathrm{i}\int{d^{2}x\left(\psi^{\ast}_L\left(\mathrm{i}\partial_-+A^L_-\right)\psi_L+\psi^{\ast}_R\left(\mathrm{i}\partial_++A^R_+\right)\psi_R\right)}}}.
\label{8}
\end{equation}

The classical action is invariant under the chiral gauge transformations
\begin{subequations}
\begin{eqnarray}
&&\psi_L\rightarrow g_L(x)\psi_L~~~\text{and}~~~A^L_-\rightarrow g_L(x)A^L_-g^{-1}_L(x)-i\partial_-g_L(x)g^{-1}_L(x)\label{9};\\
&&\psi_R\rightarrow g_R(x)\psi_R~~~\text{and}~~~A^R_+\rightarrow g_R(x)A^R_+g^{-1}_R(x)-i\partial_+g_R(x)g^{-1}_R(x).\label{10}
\end{eqnarray}
\end{subequations}
At the quantum level, however, this invariance is broken due to the noninvariance of the fermionic path-integral measure leading to the non-Abelian chiral anomaly
\begin{equation}
W[A^{\theta_L}_L,A^{\theta_R}_R]-W[A_L,A_R]\propto\int{d^2x\text{Tr}\left(\partial_+\theta_LA^L_-+\partial_-\theta_LA^L_+-\partial_-\theta_RA^R_+-\partial_+\theta_RA^R_-\right)},\label{13}
\end{equation}
where $A^\theta$ is the gauge transformation of $A$ by an infinitesimal gauge parameter $\theta$.

It is important to know the gauge redundancies of the quantum theory in order to identify the corresponding physical degrees of freedom. If we constrain the gauge fields to satisfy $A^R=A^L$ and consequently $\theta_L=\theta_R$ we get $\delta W=0$ in (\ref{13}). Such a vector gauge coupling implemented on a system of initially free fermions could only describe non-chiral CFT \cite{Cabra}. Since our goal is to construct chiral coset CFT, we will consider instead independent $A^R$ and $A^L$ gauge fields with the corresponding anomalous gauge symmetries (\ref{9}) and (\ref{10}). Firstly, one should notice that in this more general case we have a residual chiral gauge symmetry with $\theta_L=\theta_L(x^-)$ and $\theta_R=\theta_R(x^+)$, with $x^{\pm}=\frac{1}{2}(t\pm x)$, which is not anomalous. This is special for two dimensions and it was first noticed in \cite{Tye}. In that work the authors start with the WZW model with global $G\times G$ invariance and gauge different subgroups $H_L$ and $H_R$ of the $G$ sectors. It is then argued that after fixing the chiral gauge symmetry one would obtain a $G/H_L\times G/H_R$ coset model.
However, this chiral symmetry is not a local symmetry in the sense that a local parameter should depend on both light-cone coordinates. Therefore one cannot use this symmetry to suppress dynamical degrees of freedom as is usual for gauge symmetries\footnote{Since we only have two gauge parameters with half of the coordinates, we can only fix the gauge along two curves in the plane. If we then factor out the group volumes, we obtain a measure in the partition function that differs from the initial gauge measure by a null measure set.}. As pointed in \cite{Tseytlin} the central charge of the {\it chiral gauged WZW model} of \cite{Tye} corresponds to a chiral CFT but cannot describe the desired coset structure $G/H_L\times G/H_R$.

To proceed with the identification of the possible CFTs described by (\ref{7}), we integrate out the fermionic fields in (\ref{8}) to get
\begin{equation}
e^{\mathrm{i}W[A^L,A^R]}=\det(D_-)\det(D_+).\label{14}
\end{equation}
As usual, it is convenient to parametrize the gauge fields in terms of group-valued fields:
\begin{subequations}
\begin{eqnarray}
A^R_+&=&\mathrm{i}\partial_+RR^{-1},\ \ \ \ R\in H_R,\label{15}\\
A^L_-&=&-\mathrm{i}L^{-1}\partial_-L,\ \ \ \ L\in H_L.\label{16}
\end{eqnarray}
\end{subequations}
The determinants of the covariant derivatives in (\ref{14}) are calculated for the vector gauge coupling, $N^L_f=N^R_f=N_f$ and $A_L=A_R$, in \cite{Polyakov1,Polyakov2}. In this case the calculation is most easily performed choosing the gauge $A_-=0$, which yields
\begin{equation}
\det(\slashed D)=\det(\partial_-)\det(D_+)=e^{-\mathrm{i}N_fS_{WZW}[R]},\label{62}
\end{equation}
up to field independent normalization factors, with $S_{WZW}[R]$ being the Wess-Zumino-Witten action for the group-valued field $h$:
\begin{equation}
S_{WZW}[R]=\frac{1}{8\pi}\int{d^2x\text{Tr}\left(\partial_\mu R^{-1}\partial^\mu R\right)}+\frac{1}{12\pi}\int{d^3x\epsilon^{\mu\nu\rho}\text{Tr}\left(R^{-1}\partial_\mu RR^{-1}\partial_\nu RR^{-1}\partial_\rho R\right)}.\label{19}
\end{equation}
The WZW action is invariant under the affine Kac-Moody transformation $R(x^+,x^-)\rightarrow \Omega_L(x^-)R(x^+,x^-)\Omega^{-1}_R(x^+)$, with $\Omega_L$ and $\Omega_R\in G$, which leads to the holomorphic conservation of the chiral currents: $\partial_+J_L=0$ and $\partial_-J_R=0$, with $J_L=-\frac{\mathrm{i}}{4\pi}\partial_+RR^{-1}$ and $J_R=\frac{\mathrm{i}}{4\pi}R^{-1}\partial_-R$.

It is interesting to notice how the calculation (\ref{62}) of the fermionic determinant is equivalent to non-Abelian bosonization. We start by considering a system of $N$ Dirac free fermions, which has a global $U(N)$ symmetry, coupled to a external gauge field valued in the Lie algebra of a subgroup $SU(N_c)$ of $U(N)$. To calculate the effective action of this system via bosonization rules, it is sufficient to consider the subgroup $G=SU(N_f)\times SU(N_c)\times U(1)$ of $U(N)$ \cite{witten}, with $N=N_fN_c$ \cite{Frishman}. Let $u$, $h$, and $\phi$ be group-valued fields in $SU(N_f)$, $SU(N_c)$, and $U(1)$, respectively. Then, the $G$-valued field $g$ can be written as $g=uh\phi$. The relevant non-Abelian bosonization rules are \cite{witten2,Frishman}:
\begin{subequations}
\begin{eqnarray}
\bar{\psi}^{i\alpha}\mathrm{i}\slashed\partial\psi^{i\alpha}&\rightarrow&S[uh\phi]=N_cS_{WZW}[u]+N_fS_{WZW}[h]+NS[\phi],\\
\psi^{i\alpha\ast}_LT^A_{\alpha\beta}\psi^{i\beta}_L&\rightarrow&-\frac{\mathrm{i}}{4\pi}N_f\left(\partial_+hh^{-1}\right)^{\alpha\beta}T^A_{\beta\alpha},\\
\psi^{i\alpha\ast}_RT^A_{\alpha\beta}\psi^{i\beta}_R&\rightarrow&\frac{\mathrm{i}}{4\pi}N_f\left(h^{-1}\partial_-h\right)^{\alpha\beta}T^A_{\beta\alpha},
\end{eqnarray}
\end{subequations}
with $i=1,\ldots,N_f$; $\alpha$, $\beta=1,\ldots,N_c$; $A=1,\ldots,N^2_c-1$, and $T^A$ are the generators of $SU(N_c)$. Using these rules, we have
\begin{eqnarray}
\det(\partial_-)\det(D_+)&=&\int{\mathcal{D}\bar{\psi}\mathcal{D}\psi e^{\mathrm{i}\int{d^2x\left(\bar{\psi}\mathrm{i}\slashed\partial\psi+A^{\alpha\beta}_+\psi^{i\alpha\ast}_R\psi^{i\beta}_R\right)}}}\nonumber\\
&=&\int{\mathcal{D}u\mathcal{D}\phi e^{\mathrm{i}\left(N_cS_{WZW}[u]+NS_{WZW}[\phi]\right)}}\int{\mathcal{D}he^{\mathrm{i}N_f\left(S_{WZW}[h]-\frac{1}{4\pi}\int{d^2x\text{Tr}\left(\partial_+RR^{-1}h^{-1}\partial_-h\right)}\right)}}\nonumber\\
&=&\text{const}\times e^{-\mathrm{i}N_fS_{WZW}[R]}\int{\mathcal{D}he^{\mathrm{i}N_fS_{WZW}[hR]}}\nonumber\\
&=&e^{-\mathrm{i}N_fS_{WZW}[R]}\times\text{const}.\label{61}
\end{eqnarray}
From the second to third line we have written the integral over the fields that do not couple to the external gauge field as a constant. Also we used the Polyakov-Wiegmann identity
\begin{eqnarray}
S_{WZW}[hR]=S_{WZW}[h]+S_{WZW}[R]-\frac{1}{4\pi}\int{d^2x\text{Tr}\left(\partial_+RR^{-1}h^{-1}\partial_-h\right)}\label{63}
\end{eqnarray}
and the parametrization $A_+=\mathrm{i}\partial_+RR^{-1}$. To absorb the integral in the third line in the normalization constant, we just need to perform the change of variables $hR\rightarrow h$ and use the invariance of the integration measure. We then regain the expression (\ref{62}) for the fermionic determinant.

We can even be more general and add a local counterterm $\alpha\int{d^2x\text{Tr}\left(A_+A_-\right)}$ in the bosonic action (\ref{61}) \cite{Jackiw}. This freedom reflects the arbitrariness in the choice of the regularization scheme to define the fermionic measure. However, in this case it is natural to choose $\alpha=0$, which amounts to regularize the system preserving the original vector gauge symmetry. In the more general context when we consider different gauge subgroups we cannot choose a gauge invariant regularization due to the chiral anomaly and it is more natural to let $\alpha$ be arbitrary.

We will see that we can further generalize the bosonization procedure to have independent left and right sectors, called chiral bosonization \cite{Sonnenschein}, which is the appropriate formalism to construct the chiral coset structure. To this end, we consider the separate calculation of the two determinants $\det(D_-)$ and $\det(D_+)$. As is well known, since the operators $D_\pm$ do not have a well-defined eigenvalue problem, to make sense of their determinants one needs to supplement the effective actions $W[A_\pm]$ with the free part of the missing fermion chirality \cite{gaume}. Then for each determinant the calculation turns into the same performed in \cite{Polyakov1,Polyakov2} with the caveat that to recover the particular case with only one fermion chirality coupled to an external gauge field we should impose a constraint on the WZW action to allow only one bosonic chirality to be dynamical \cite{Sonnenschein}.

Instead of the WZW model (\ref{19}), for the left-right decoupling analysis, it is convenient to consider a chiral-split form that generalizes the WZW action
\begin{equation}
S_{LR}[g_R, g_L]=S^{+}_{ch}[g_R]+S^{-}_{ch}[g_L],\label{20}
\end{equation}
where the chiral pieces are given by
\begin{eqnarray}
S^{\pm}_{ch}[g]&=&\mp\frac{1}{4\pi}\int{d^2x\text{Tr}\left(\partial_x g^{-1}\partial_{\pm} g\right)}+\Gamma[g],\label{21}
\end{eqnarray}
with $\Gamma[g]$ being the Wess-Zumino term corresponding to the second term in (\ref{19}). In (\ref{20}), each chiral action possesses only half of the Kac-Moody symmetry, $u_R\rightarrow g_R\Omega^{-1}_R(x^+)$ and $g_L\rightarrow\Omega_L(x^-)g_L$, compared with $S_{WZW}$. These symmetries lead to the holomorphic conservations: $\partial_+J^{(-)}_L=0$ and $\partial_-J^{(+)}_R=0$, with $J^{(-)}_L=-\frac{\mathrm{i}}{2\pi}\partial_xg_Lg_L^{-1}$ and $J^{(+)}_R=-\frac{\mathrm{i}}{2\pi}g^{-1}_R\partial_xg_R$. The superindexes $\pm$ in the currents $J^{(\pm)}$ refer to the corresponding actions $S^{(\pm)}_{ch}$. We also have the conservations $\partial_\mu J^{(+)\mu}_L=0$ and $\partial_\mu J^{(-)\mu}_R=0$ in the non-conformal sectors of the chiral actions. The equations of motion that follow from $S^{-}_{ch}(g_L)$ ($S^{+}_{ch}(g_R)$) render the on-shell condition $g_L(x)=\tilde{g}_L(x^+)A(t)$ ($g_R(x)=B(t)\tilde{g}_R(x^-)$)  and the equivalence of $S_{LR}$ with the original WZW model is achieved upon the constraint that the non-conformal modes in the chiral actions cancel each other ($A=B^{-1}$) \cite{Wotzasek}. Therefore, for the purpose of counting the conformal degrees of freedom we can use the Knizhnik-Zamolodchikov (KZ) formula to compute the central charge of the sector in the chiral action that exhibits the KM symmetry.

We argue that we can identify the chiral determinant $\det(D_{+})$ ($\det(D_{-})$) with the chiral WZW model $S^{+}_{ch}(g_{R})$ ($S^{-}_{ch}(g_{L})$). To support this identification we use the bosonization rules to express chiral fermions in terms of the chiral WZW model \cite{Sonnenschein}. Again, we start by considering a system of $N$ Dirac free fermions, which has a global $U(N)_R\times U(N)_L$ symmetry. But, now we allow independent couplings of external gauge fields to the left and right sectors. To understand the general picture it is enough to consider the subgroups $SU(N^L_c)$ and $SU(N^R_c)$ of $U(N)_L$ and $U(N)_R$, respectively. We want to calculate the generating functional for Green functions involving the fermionic currents associated to these subgroups. Then, generalizing the above discussion, the calculation of the effective action can be performed by using bosonization rules focusing on the group structure $G_L\times G_R=\left(SU(N^L_f)\times SU(N^L_c)\times U(1)_L\right)\times\left(SU(N^R_f)\times SU(N^R_c)\times U(1)_R\right)$, with $N^L_fN^L_c=N^R_fN^R_c=N$. To keep the chiral sectors independent we use non-Abelian bosonization of chiral fermions. Taking $g_L$ ($g_R$) $\in$ $G_L$ ($G_R$) as $g_L=u_Lh_L\phi_L$ ($g_R=u_Rh_R\phi_R$) with $u_L$ ($u_R$), $h_L$ ($h_R$), and $\phi_L$ ($\phi_R$) belonging to $SU(N^L_f)$ ($SU(N^R_f)$), $SU(N^L_c)$ ($SU(N^R_c)$), and $U(1)_L$ ($U(1)_R$), respectively, and  following \cite{Sonnenschein}, we have 
\begin{subequations}
\begin{eqnarray}
\psi^{\ast i\alpha}_L\mathrm{i}\partial_-\psi^{i\alpha}_L&\rightarrow& S^{-}_{ch}[u_Lh_L\phi_L]=N^L_cS^{-}_{ch}[u_L]+N^L_fS^{-}_{ch}[h_L]+NS^{-}_{ch}[\phi_L],\\
\psi^{\ast i^\prime\alpha^\prime}_R\mathrm{i}\partial_+\psi^{i^\prime\alpha^\prime}_R&\rightarrow& S^{+}_{ch}[u_Rh_R\phi_R]=N^R_cS^{+}_{ch}[u_R]+N^R_fS^{+}_{ch}[h_R]+NS^{+}_{ch}[\phi_R]\\
\psi^{\ast i\alpha}_LT^{(L)A}_{\alpha\beta}\psi^{i\beta}_L&\rightarrow& -\frac{\mathrm{i}}{2\pi}N^L_f\left(\partial_xh_Lh^{-1}_L\right)^{\alpha\beta}T^{(L)A}_{\beta\alpha},\\
\psi^{\ast i^\prime\alpha^\prime}_RT^{(R)A^\prime}_{\alpha^\prime\beta^\prime}\psi^{i^\prime\beta^\prime}_R&\rightarrow& -\frac{\mathrm{i}}{2\pi}N^R_f\left(h^{-1}_R\partial_xh_R\right)^{\alpha^\prime\beta^\prime}T^{(R)A^\prime}_{\beta^\prime\alpha^\prime},
\end{eqnarray}
\end{subequations}
which are adequate for the Left-Right regularization scheme. We then obtain
\begin{eqnarray}
\det(D_+)&=&\int{\mathcal{D}\psi^\ast_R\mathcal{D}\psi_Re^{\mathrm{i}\int{d^{2}x\psi^{\ast}_R\left(\mathrm{i}\partial_++A^R_{+}\right)\psi_R}}}\nonumber\\
&=&\text{const}\times\int{\mathcal{D}h_Re^{\mathrm{i}N^{R}_f\left[S^{+}_{ch}[h_R]-\frac{\mathrm{i}}{2\pi}\int{d^2x\text{Tr}\left(A^R_{+}h^{-1}_R\partial_xh_R\right)}-\frac{\alpha}{4\pi}\int{d^2x\text{Tr}\left(A^R_{+}A^R_x\right)}\right]}},\label{64}
\end{eqnarray}
and, similarly
\begin{equation}
\det(D_-)=\text{const}\times\int{\mathcal{D}h_Le^{\mathrm{i}N^{L}_f\left[S^{-}_{ch}[h_L]-\frac{\mathrm{i}}{2\pi}\int{d^2x\text{Tr}\left(A^L_{-}\partial_xh_Lh^{-1}_L\right)}+\frac{\alpha}{4\pi}\int{d^2x\text{Tr}\left(A^L_{-}A^L_x\right)}\right]}}.\label{65}
\end{equation}
As before, we have put the integration over the decoupled fields $u$ and $\phi$ in the normalization constant. The last terms in the above bosonic actions with the parameter $\alpha$ are possible counter-terms we can add and adjust according to the chosen regularization scheme to treat the fermionic measures.

Using the parametrizations (\ref{15}) and (\ref{16}) and the chiral Polyakov-Wiegmann (PW) identities,
\begin{subequations}
\begin{equation}
S^+_{ch}[hR]=S^+_{ch}[h]+S^+_{ch}[R]+\frac{1}{2\pi}\int{d^2x\text{Tr}\left(h^{-1}\partial_xh\partial_+RR^{-1}\right)},\label{66}
\end{equation}
\begin{equation}
S^-_{ch}[hL]=S^-_{ch}[h]+S^-_{ch}[L]-\frac{1}{2\pi}\int{d^2x\text{Tr}\left(\partial_xhh^{-1}L^{-1}\partial_-L\right)},\label{67}
\end{equation}
\end{subequations}
we can rewrite the bosonic integrals (\ref{64}) and (\ref{65}) as
\begin{equation}
\int{\mathcal{D}h_Le^{\mathrm{i}N^{L}_f\left(S^{-}_{ch}[h_LL]-S^{-}_{ch}[L]+\frac{\alpha}{4\pi}\int{d^2x\text{Tr}\left(A^L_{-}A^L_x\right)}\right)}}=\text{const}\times e^{-\mathrm{i}N^{L}_f\left(S^{-}_{ch}[L]-\frac{\alpha}{4\pi}\int{d^2x\text{Tr}\left(A^L_{-}A^L_x\right)}\right)},\label{23}
\end{equation}
and
\begin{equation}
\int{\mathcal{D}h_Re^{\mathrm{i}N^{R}_f\left(S^{+}_{ch}[h_RR]-S^{+}_{ch}[R]-\frac{\alpha}{4\pi}\int{d^2x\text{Tr}\left(A^R_{+}A^R_x\right)}\right)}}=\text{const}\times e^{-\mathrm{i}N^{R}_f\left(S^{+}_{ch}[R]+\frac{\alpha}{4\pi}\int{d^2x\text{Tr}\left(A^R_{+}A^R_x\right)}\right)},\label{24}
\end{equation}
since we can make the shift $h_LL\rightarrow h_L$ ($h_RR\rightarrow h_R$). Choosing the normalization factor to give $\det(D_{\pm})=\det(\partial_{\pm})$ when $A_{\pm}=0$, we then obtain for $e^{\mathrm{i}W[A^L,A^R]}$:
\begin{eqnarray}
\det(D_-)\det(D_+)&=&\left(\det\partial_-\right)\left(\det\partial_+\right)\exp{\left[-\mathrm{i}\left(N^R_fS^+_{ch}[R]+N^L_fS^{-}_{ch}[L]\right.\right.}\nonumber\\
&+&\left.\left.\frac{\alpha}{8\pi}\int{d^2x\text{Tr}\left(N^R_f\left(A^R_+\right)^2+N^L_f\left(A^L_-\right)^2-N^R_fA^R_+A^R_--N^L_fA^L_+A^L_-\right)}\right)\right].\label{25}
\end{eqnarray}

It is interesting to check the consistency of our analysis by examining the case $N^L_f=N^R_f$ and $A^R=A^L=A$. In this situation the fermionic system has an increased non-anomalous gauge symmetry compared with the case with $N^L_f\neq N^R_f$. The value $\alpha=1$ for the regularization-dependent parameter is the adequate one in this case, since we have the algebraic identities $S^{\pm}_{ch}+\frac{1}{8\pi}\left(A_{\pm}\right)^2=S_{WZW}$ and the effective action in (\ref{25}) amounts to $S[L]+S[R]-\frac{1}{4\pi}\int{d^2x\text{Tr}\left(A_+A_-\right)}$. This, in turn, is nothing else than $S[LR]$ by using the PW identity. The same vector gauge symmetry, $L^\prime=L\lambda$ and $R^\prime=\lambda^{-1}R$, is then recovered in terms of the effective action. In this work, on the other hand, we are mainly interested in the general case when $N^L_f\neq N^R_f$. In this situation it is convenient to choose $\alpha=0$, since we then obtain a left-right decoupling of the degrees of freedom with the conformal symmetry in the corresponding sector in agreement with the symmetries of the original fermionic model. Therefore, we will consider $\alpha=0$ henceforth. 

The two determinants of ordinary derivatives in (\ref{25}) can be exponentiated back in terms of free fermions. The partition function (\ref{7}) is obtained by further integrating this expression over the gauge fields. Changing the integration variables from $A^R$ and $A^L$ to $R$ and $L$, respectively, we get
\begin{equation}
\mathcal{D}A_R\mathcal{D}A_L=J[R]J[L]\mathcal{D}R\mathcal{D}L,
\end{equation}
with the non-trivial Jacobians given by
\begin{equation}
J[R]J[L]=\left(\det\partial_-\right)_{SU(N^L_c)}\left(\det\partial_+\right)_{SU(N^R_c)}e^{-2\mathrm{i}C_{SU(N^R_c)}S^{+}_{ch}[R]}e^{-2\mathrm{i}C_{SU(N^L_c)}S^{-}_{ch}[L]},\label{26}
\end{equation}
where $C_H$ is the quadratic Casimir invariant for the subgroup $H$ in the adjoint representation. For $SU(N)$, $C_H=N$, while for $U(1)$, we have $C_H=0$.
The two free chiral determinants $\left(\det\partial_{\pm}\right)$ can be written in terms of ghost fields $b_{\pm},c,\bar{c}$:
\begin{equation}
\left(\det\partial_-\right)_{SU(N^L_c)}\left(\det\partial_+\right)_{SU(N^R_c)}=\int{\mathcal{D}c\mathcal{D}b_+\mathcal{D}\bar{c}\mathcal{D}b_-e^{\mathrm{i}\int{d^2x\text{Tr}\left(b_+\partial_-\bar{c}+b_-\partial_+c\right)}}}.
\end{equation}
The ghosts $b_{\pm}$ have conformal weight 1, whereas $c$ and $\bar{c}$ have conformal weight 0. These ghost systems contributes negatively to the central charge of the chiral sectors as $c^{L/R}_{ghost}=-2\text{dim}(\hat{H}_{L/R})$. Using these results we can rewrite the partition function in terms of free fermions, ghost systems and WZW models. We can then proceed and calculate the total central charge of the system as $c_T=c_{matter}+c_{ghost}+c_{WZW}$ for the both sectors. With this is mind and generalizing the construction for multiple current constraints on each sector we obtain for the total central charge of the system
\begin{subequations}
\begin{eqnarray}
c_R&=&N+\sum_i\left(-2\text{dim}\left(\hat{H}^i_R\right)+\frac{\left(N^{R(i)}_f+2C_{H^i_R}\right)\text{dim}\left(\hat{H}^i_R\right)}{N^{R(i)}_f+C_{H^i_R}}\right),\label{27}\\
c_L&=&N+\sum_i\left(-2\text{dim}\left(\hat{H}^i_L\right)+\frac{\left(N^{L(i)}_f+2C_{H^i_L}\right)\text{dim}\left(\hat{H}^i_L\right)}{N^{L(i)}_f+C_{H^i_L}}\right),\label{28}
\end{eqnarray}
\end{subequations}
where we have used the KZ formula for the central charge of the WZW model. To apply this formula for the $H=U(1)$ case, we remember that the WZW model then reduces to the free boson model. In this case, and as it is known, such a model describes a  CFT with $c=1$. Furthermore, we take $\text{dim}(u(1))=1$ and, as we already pointed, $C_{U(1)}=0$. For the $H=SU(N_c)$ case we have $\text{dim}(\hat{H})=N^2_c-1$, $C_H=N_c$ and $N^{L/R(i)}_f=N/N^{L/R(i)}_c$.

We can compare our procedure to get the results (\ref{27}) and (\ref{28}) with that of Ref. \cite{Tye}. The reason why we obtain in our construction the coset structure $G/H_L\times G/H_R$ is related to the chiral bosonization prescription we adopted to treat the chiral determinants. Therefore, the degrees of freedom that should be suppressed in the approach of \cite{Tye} in order to produce the structure $G/H_L\times G/H_R$ do not appear from the onset since equations (\ref{25}) and (\ref{26}) involve only the chiral pieces of the WZ actions.

\begin{figure}[!h]
\centering
\includegraphics[scale=0.7]{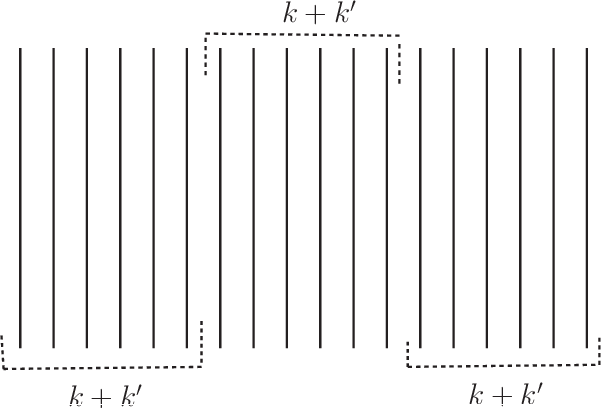}
\caption{System of wires decomposed into bundles. The picture shows three bundles constituted of $k+k'=6$ wires.}
\label{Wiresdec}
\end{figure}


\section{Bulk-Edge Correspondence}\label{BulkEdge}

In the last section we saw that the prescriptions (\ref{64}) and (\ref{65}) for the chiral determinants provide the right structure to produce independent cosets $G/H_R$ and $G/H_L$ for right and left sectors. This is the desirable feature that is useful to describe the edge of a chiral topological phase. 

Now we employ the previous formalism to make one important connection between the chiral constrained fermions and the bulk theory, which is given in terms of a topological effective field theory. This can be reached by exploring the bulk-edge correspondence \cite{Wen,Rossini,Seiberg}. The key point is the identification of the chiral fermions in terms of a chiral gauged WZW model, which in turn can be obtained from a Chern-Simons-type theory defined on a manifold with a boundary. In our work, one can make a direct construction by starting with a CS model $S_{CS}[B]$, with $B$ being a gauge field valued in the Lie-algebra of a group $G$,
\begin{eqnarray}
S_{CS}[B]&=&\frac{k}{4\pi}\int_{\Omega}{\text{Tr}\left(BdB+\frac{2}{3}B^3\right)}.
\label{29}
\end{eqnarray}
We then impose the boundary conditions $B_t-B_x-\pi_H(B_t-B_x)=0$ and $\pi_H(B_i)=0$, with $\pi_H$ being the orthogonal projection in Killing metric of the Lie-algebra of $G$ onto the Lie-algebra of the subgroup $H$ and supposing the boundary is at $y=0$. We explore the general coordinate invariance of this action to work in a coordinate system $(t^\prime,~x^\prime,~y^\prime)$ defined by
\begin{eqnarray}
t^\prime&=&t,\\
x^\prime&=&t+x,\\
y^\prime&=&y,
\end{eqnarray}
where the boundary conditions reduce to $B^\prime_0-\pi_H(B^\prime_0)=0$ and $\pi_H(B^\prime_i)=0$. Decomposing the exterior derivative $d=dt^\prime\frac{\partial}{\partial t^\prime} +\tilde{d}^\prime$ and the gauge field $B=B^\prime_t+\tilde{B}^\prime$ into space and time components and using this alternative boundary condition, we have
\begin{eqnarray}
S_{CS}&=&-\frac{1}{4\pi}\int_{\Omega}{\text{Tr}\left(\tilde{B}^\prime\frac{\partial}{\partial t^\prime}\tilde{B}^\prime\right)dt^\prime}-\frac{1}{2\pi}\int_{\Omega}{\text{Tr}\left(B^\prime_t\left(\tilde{d}^\prime\tilde{B}^\prime+\tilde{B}^{\prime2}\right)\right)}.
\end{eqnarray}
The integration over $B^\prime_t$ imposes the constraint $\tilde{F}^\prime=0$, which implies that the space components are pure gauge, $\tilde{B}^\prime=-\mathrm{i}h_R^{-1}\tilde{d}^\prime h_R$, with $h_R\in G$. Plugging this solution back into the action, integrating by parts, and expressing in terms of the old coordinates, we get
\begin{eqnarray}
S_{CS}&=&\frac{1}{4\pi}\int_{\partial\Omega}{\text{Tr}\left(h_R^{-1}\partial_+h_Rh_R^{-1}\partial_xh_R\right)dxdt}\nonumber\\
&+&\frac{1}{12\pi}\int_{\Omega}{\text{Tr}\left(h_R^{-1}dh_R\right)^3}-\frac{\mathrm{i}}{2\pi}\int{d^2x\text{Tr}\left(\lambda h_R^{-1}\partial_xh_R\right)},
\end{eqnarray}
with $\lambda$ being a Lagrange multiplier valued in the Lie-algebra of a subgroup $H$ of $G$ that imposes the boundary condition $\pi_H\left(B_i\right)=0$. Identifying $\lambda$ with a gauge field $A^R_+$ of the subgroup $H$, we obtain the gauged chiral WZW model in the Left-Right regularization scheme. Therefore, we can consider the action (\ref{29}) with the associated boundary conditions as the bulk theory corresponding to our coset CFT. This construction also provides an explicit link between the quantum wires construction for chiral CFT models with the effective CS bulk theory. 

In \cite{Rossini}, the bulk-edge connection is made with the edge theory described in terms of the chiral gauged WZW model but in the Vector-Axial regularization scheme. To get a coset theory they start with a Chern-Simons model with a $G\times H$ gauge structure of the type $S_{CS}[A]-S_{CS}[B]$ with $A$ and $B$ being the gauge fields valued in the Lie-algebras of $G$ and $H$, respectively. Due to the minus sign in the $H$ sector, the conformal structure corresponds to a holomorphic CFT in the $G$-sector and an anti-holomorphic CFT in the $H$-sector. To realize a coset $G/H$ conformal structure, with $H$ being a subgroup of $G$, one imposes the boundary conditions $A_0-A_x-\pi_H(A_0-A_x)=0$ and $\pi_H(A_i)=B_i$. The boundary theory is given by the chiral gauged WZW model in the Vector-Axial regularization scheme:
\begin{equation}
\int Dh_RDA_-DA_xe^{iS_{CG}^{VA(+)}[h_R,A_+,A_x]},
\end{equation}
with
\begin{eqnarray}
S_{CG}^{VA(+)}[h_R,A_+,A_x]&=&S^+_{ch}[h_R]+\frac{1}{2\pi}\int d^2x\text{Tr}\left(h^{-1}_RA_xh_RA_+-iA_+h^{-1}_R\partial_xh_R\right.\nonumber\\&+&\left. iA_x\partial_+h_Rh^{-1}_R-A_+A_x\right).
\end{eqnarray}
To compare with the construction of \cite{Rossini}, we use the identities
\begin{equation}
S_{CG}^{LR(+)}[h_R,A_+]=S_{ch}^{+}[h_RR]-S_{ch}^{+}[R]\label{78}
\end{equation}
and
\begin{equation}
S_{CG}^{VA(+)}[h_R,A_+,A_x]=S_{ch}^{+}[Xh_RR]-S_{ch}^{+}[XR],\label{79}
\end{equation}
where $A_+=\mathrm{i}\partial_+RR^{-1}$ and $A_x=-\mathrm{i}X^{-1}\partial_xX$, with $R$ and $X\in H$. The regularization in the $VA$ scheme preserves the conservation of the vector current, $D_\mu J^\mu=0$. This is compatible with the presence of the gauge vector symmetry $R^\prime=\lambda R$ and $X^\prime=X\lambda^{-1}$, which can be verified explicitly by a change of variables $h^\prime_R=\lambda h_R\lambda^{-1}$ and the invariance of the group measure. Therefore we can use the gauge freedom to fix the gauge $X=1$ in (\ref{79}), showing the equivalence between the two effective actions. Analogously, we can implement the Stueckelberg procedure in (\ref{78}) by introducing a new dynamical variable with a convenient gauge symmetry. To this end we make the same gauge transformation as before, but now in (\ref{78}). After the change of variables $h^\prime_R=\lambda h_R\lambda^{-1}$ we get $S_{ch}^{+}[\lambda h_RR]-S_{ch}^{+}[\lambda R]$. Then, by promoting $\lambda$ to a dynamical variable $X$, which transforms as $X^\prime=X\lambda^{-1}$, we regain the action (\ref{79}). We can also verify the equivalence between the constructions already at the level of the bulk CS theories. In fact, from the bulk theory $S_{CS}[A]-S_{CS}[B]$ and the boundary conditions $A_0-A_x-\pi_H(A_0-A_x)=0$ and $\pi_H(A_i)=B_i$, we recognize a residual gauge invariance $A^\prime_i=\lambda A_i\lambda^{-1}-i\partial_i\lambda\lambda^{-1}$ and $B^\prime_i=\lambda B_i\lambda^{-1}-i\partial_i\lambda\lambda^{-1}$. Since the fields $A_i$ and $B_i$ become pure gauges after the $A_0$ and $B_0$ integrations, the residual gauge invariance is sufficient to eliminate $B_i$ completely. Therefore, in this gauge, the bulk theory is equivalent to $S_{CS}[A]$ with the boundary conditions $A_0-A_x-\pi_H(A_0-A_x)=0$ and $\pi_H(A_i)=0$.

\section{Description of Topological Phases\label{S3}}

In this section we show how the previous formalism can be used to describe the chiral edge states of a class of topological phases firstly proposed in  Ref. \cite{Huang}. As discussed previously, the effect of constraining fermionic currents is to project out of the spectrum the corresponding degrees of freedom. In the quantum wires context this is equivalent to gap the conformal degrees of freedom associated to the currents.  

For the class of phases we are interested in this section, it is convenient to start from a specific arrangement of the wires. Thus, by following \cite{Huang} we split the set of quantum wires into $B$ bundles of $k+k^{\prime}$ quantum wires, as shown in Fig. \ref{Wiresdec}. The total number of wires is then $N=B (k+k')$. Consider each wire supporting a pair of Dirac fermions $\psi_{R/L,\sigma,I}$, with $\sigma=1,2$ being the spin and $I=1,...,N$ being the wire index. We adapt the index structure to the bundle arrangement as $\psi_{R/L,\sigma,I}\rightarrow \psi_{R/L,\sigma,i}^m~\text{and}~\psi_{R/L,\sigma,i'}^m$, with $m=1,\ldots,B$ specifying the bundles, $i=1\ldots,k$ and $i^{\prime}=1\ldots,k^\prime$, specifying the wires inside each bundle. The free Lagrangian can then be written as
\begin{equation}
\mathcal{L}_0=\sum_{m,i,\sigma}\mathrm{i}\left(\psi_{R,\sigma,i}^{m\ast}\partial_+\psi^{m}_{R,\sigma,i}+\psi_{L,\sigma,i}^{m\ast}\partial_-\psi^{m}_{L,\sigma,i}\right)
+\sum_{m,i^\prime,\sigma}\mathrm{i}\left(\psi_{R,\sigma,i^{\prime}}^{m\ast}\partial_+\psi^{m}_{R,\sigma,i^{\prime}}+\psi_{L,\sigma,i^{\prime}}^{m\ast}\partial_-\psi^{m}_{L,\sigma,i^{\prime}}\right).
\label{30}
\end{equation}
This Lagrangian describes a free conformal field theory with total central charge $c=2N$. It is invariant under time-reversal and under transformations of $G=U(2N)_R\times U(2N)_L$. 

For each bundle we have the subgroup of symmetry $U(2k)\times U(2k^\prime)$ for the left and right sectors. 
Let us consider the symmetry group $U(2k)$ of the set of $k$ wires inside a bundle. The corresponding algebra can be decomposed as
\begin{equation}
u(2k)\supset u(1)\oplus su(2)_k\oplus su(k)_2.
\label{31a}
\end{equation}
The strategy is to introduce interactions in order to gap some of these pieces. There are two classes of local interactions we can add: interactions involving wires of the same bundle (intra-bundle) and interactions involving wires of neighboring bundles (inter-bundle). Both type are required to produce a 2+1 dimensional gapped topological phase. Appendix \ref{AA} contains a detailed discussion of the form of such interactions. 

\begin{figure}[!h]
\centering
\includegraphics[scale=0.7]{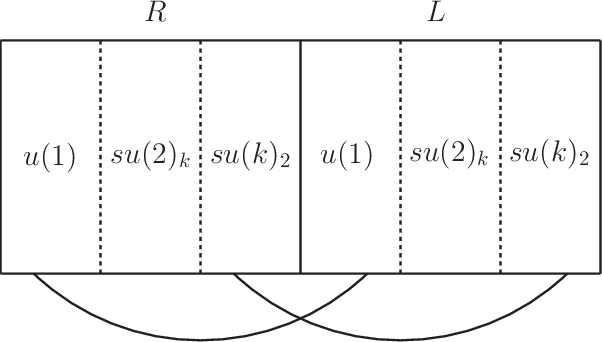}
\caption{Intra-bundle interactions that provide a gap for $U(1)$ and $SU(k)$ parts. The letters $R$ and $L$ indicate right and left sectors. A similar structure is assumed for the set of $k'$ wires omitted in the figure.}
\label{groupdec}
\end{figure}

First, we consider the effect of providing a gap to the $U(1)$ and $SU(k)$ parts. Then the gapless Hilbert space is obtained by imposing the constraints
\begin{equation}
J_{R/L}^{U(1)}|\text{phys}\rangle =0 ~~~\text{and}~~~J_{R/L}^{SU(k)}|\text{phys}\rangle =0.
\label{35}
\end{equation}
The intra-bundle interactions are depicted in Fig. \ref{groupdec} and realize the non-chiral coset structure 
\begin{equation}
\frac{u(2k)}{u(1)\oplus su(k)_2}=su(2)_k,
\label{36}
\end{equation}
since left and right sectors are treated on equal foot. A similar structure is assumed for the set of $k'$ wires. 
The central charge can be immediately computed with expressions (\ref{27}) and (\ref{28}): 
\begin{equation}
c_{R}=\sum_{m=1}^B c_R^m=c_L=\sum_{m=1}^Bc_L^m,
\label{37}
\end{equation}
where the central charge $c_{R/L}^m$ is the same for all bundles:
\begin{equation}
c_{R/L}^m=2(k+k')-1-1-\frac{2(k^2-1)}{k+2}-\frac{2(k^{\prime 2}-1)}{k'+2}.
\label{38}
\end{equation}

\begin{figure}[!h]
\centering
\includegraphics[scale=0.7]{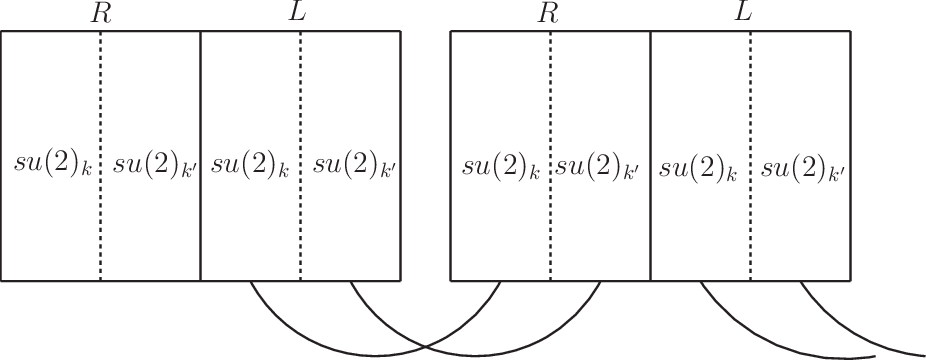}
\caption{The right sector of the first bundle contains gapless degrees of freedom, while the second bundle is fully gapped. It represents a bundle of the bulk.}
\label{su2}
\end{figure}

Until this point, we have not yet produced a 2+1 dimensional gapped phase since there are no interactions between neighboring bundles. 
Indeed, the resulting phase is a 1+1 dimensional critical phase, since there are still gapless sectors in all the bundles. The next step is to introduce interactions between neighboring bundles. Such interactions are chosen to gap the $SU(2)$ pieces and are represented in Fig. \ref{su2}. These interactions fully gap the bundles of the bulk while some sectors for the bundles of the ends are left gapless.

The second bundle of Fig. \ref{su2} is a typical bundle of the bulk that is fully gapped, i.e., both right and left central charge vanish:
\begin{equation}
c_R^m=c_L^m=2(k+k')-1-1-\frac{2(k^2-1)}{k+2}-\frac{2(k^{\prime 2}-1)}{k'+2}-\frac{3k}{k+2}-\frac{3k'}{k'+2}=0,~~~m=2,...,N-1.
\label{41}
\end{equation}
For the first bundle, however, after the imposition of the constraint 
\begin{equation}
J^{SU(2)}_L\left|\text{phys}\right>=0
\end{equation}
it remains gapless degrees of freedom associated to the right sector of the algebra 
$su(2)_k\oplus su(2)_{k'}$, which gives
\begin{equation}
c_R^{m=1}=\frac{3k}{k+2}+\frac{3k'}{k'+2}~~~\text{and}~~~ c_L^{m=1}=0.
\label{42}
\end{equation}
The last bundle contains a similar gapless content with $R\leftrightarrow L$. The whole system realizes a stable topological phase once the bulk is fully gapped while the edges contain chiral gapless modes. The phase can be Abelian (integer central charge) or non-Abelian (non-integer central charge) depending on the values of $k$ and $k'$.  

Before closing this section it is interesting to discuss the realization of topological phases whose edge states are given in terms of minimal and superconformal models. These are obtained when the diagonal subgroup $su(2)_{k+k'}$ is gapped inside each bundle (intra-bundle). Since the bulk is already completely gapped, these new interactions should only provide a gap for the bundles of the edges without destroying the bulk topological order, which is the scenario  conjectured in \cite{Huang}. 
The form of the interactions are also shown in the Appendix. The central charge of the edge CFT corresponds to the coset structure 
\begin{equation}
su(2)_k\oplus su(2)_{k^\prime}/su(2)_{k+k^\prime},\label{e43}
\end{equation}
which contains important series of conformal models. If the topological phase is stabilized by means of the mechanism proposed in \cite{Huang}, the CFT in the first bundle, for example, is realized by the constrained fermionic partition function
\begin{equation}
Z_{edge}=\int{\mathcal{D}\psi_L\mathcal{D}\psi_R\delta(J^{u(1)}_{L/R})\delta(J^{u^{\prime}(1)}_{L/R})\delta(J^{su(k)_2}_{L/R})\delta(J^{su(k^\prime)_2}_{L/R})\delta(J^{su(2)_k}_L)\delta(J^{su(2)_{k^\prime}}_L)\delta(J^{su(2)_{k+k^\prime}}_R)e^{\mathrm{i}S_0}}.\label{31}
\end{equation}
Following the reasoning applied above we get a chiral structure with the boundary CFT supporting a  central charge corresponding to the coset algebra (\ref{e43}) for the left sector, whereas the right sector is fully gapped:
\begin{subequations}
\begin{eqnarray}
c^{m=1}_L&=&0,\\
c^{m=1}_R&=&1-\frac{6k^\prime}{(k+2)(k+k^\prime+2)}+\frac{2(k^\prime-1)}{k^\prime+2}.
\end{eqnarray}
\end{subequations}
Obviously, the bulk interactions should provide the same values for the central charges to the other boundary of the system, switching the roles of the left and right sectors, in order to describe a stable topological phase. For $k^\prime=1$ and $k^\prime=2$ we reproduce the series of minimal and superconformal models, respectively.

\section{Final remarks}\label{fr}

In this work we have shown how to realize certain chiral edge states of topological phases in terms of a set of constrained fermions. As discussed,  
this system is intimately connected to the quantum wires approach since they share the same basic degrees of freedom 
and in the strong coupling limit the interactions between quantum wires can be represented by constraints on 
currents. The final effect of such constraints is to reduce the number of gapless degrees of freedom in the Hilbert space. The only 
remaining gapless degrees of freedom are located at the boundaries of the system.  

While this general picture is plausible, the system of chiral constrained fermions is subtle since it involves coupling of gauge fields to chiral fermions and this has been known for a long time to rise up issues with gauge anomalies. We have discussed how to make sense of this construction in the grounds of chiral bosonization. The bosonized form is useful as it enables to correctly identify the gapless degrees of freedom at the edge by means of the computation of the central charge. Furthermore it is the key point to establish the connection with topological bulk theory via bulk-edge correspondence.     
We believe that whole analysis can be applied for the description of other chiral topological phases in 2+1 dimensions with straightforward generalization.


\section{Acknowledgments}

We acknowledge the financial support of Brazilian agencies CAPES and CNPq.


\appendix

\section{Interactions in the Quantum Wires System}\label{AA}

In this appendix we explicitly show the interactions that provide a gap for the pieces discussed in Sec. \ref{S3}.
Let us start with the intra-bundle interactions. They are chosen to fully gap the right and left sectors of the $U(1)$ and $SU(k)$ pieces. 
To avoid heavy notation we omit the bundle index $m$ in the expressions below whenever there is no risk of ambiguities. 
The $U(1)$ interaction corresponds to a generalized Umklapp process,
\begin{equation}
\mathcal{L}_{U(1)}=- g_{u(1)} \left(\prod_{i=1}^k \prod_{\sigma=1}^2\psi^{\ast}_{R,\sigma,i} \right) 
\left(\prod_{i=k}^1 \prod_{\sigma=2}^1\psi_{L,\sigma,i} \right) + \text{H. c.}.
\label{32}
\end{equation}
This interaction gaps the $U(1)$ charge sector of wires $1,...,k$ inside a bundle. The interactions that gap the $SU(k)$ sector are of the current-current type:
\begin{equation}
\mathcal{L}_{SU(k)}=-\lambda_{SU(k)}\sum_{A=1}^{k^2-1} J_R^A J_L^A,
\label{33}
\end{equation}
namely, it gaps the $SU(k)$ sector for the wires $1,...,k$ for $\lambda_{SU(k)}>0$. The $SU(k)$ currents are given by
\begin{equation}
J_{R/L}^A=\sum_{\sigma=1}^2\sum_{i,j=1}^k\psi^{\ast}_{R/L,\sigma,i}T_{ij}^A\psi_{R/L,\sigma,j},
\label{34}
\end{equation}
where $T_{ij}^A$ are the generators of $SU(k)$, with $A=1,...,k^2-1$. As shown in \cite{Huang} by computing the renormalization group beta functions, the coupling constant $\lambda_{SU(k)}$ flows to strong coupling at low energies. Of course, the same structure of interactions is assumed for the remaining $k'$ wires of the bundle. The intra-bundle interactions are represented in Fig. \ref{groupdec}. 

Next we consider inter-bundle interactions, which turn the system of wires into a two-dimensional one. The interactions that gap the conformal degrees of freedom discussed in Sec.\ref{S3} are
\begin{equation}
\mathcal{L}_{su(2)}^{inter}=-\sum_{m=1}^{N-1}\sum_{a=1}^{3}
\left( \lambda_{m}^a J_{L}^{a,m}J_{R}^{a,m+1}+\lambda_{m}^{\prime a} J_{L}^{\prime a,m}J_{R}^{\prime a,m+1}\right),
\label{39}
\end{equation}
with the $SU(2)$ currents given by
\begin{equation}
J_{R/L}^{a,m}=\sum_{\sigma,\rho=1}^2 \sum_{i=1}^{k} \psi^{m\ast}_{R/L,\sigma,i}\frac{\sigma_{\sigma\rho}^a}{2} \psi_{R/L,\rho,i}^m,
\label{40}
\end{equation}
where $\sigma^a/2$ are the $SU(2)$ generators, $a=1,2,3$. The currents $J_{R/L}^{\prime a,m}$ refer to the set of $k'$ wires. Notice that we have reinserted the bundle index since the above interactions involve two consecutive bundles. These interactions are represented in Fig.  \ref{su2}. As in the previous case, the renormalization group calculations show that $\lambda_m$ and $\lambda_{m'}$ flow to strong coupling at low energies \cite{Huang}.

Finally, we discuss the intra-bundle interactions that are responsible for producing an interesting class of edge CFT including the minimal models as well as the superconformal models. They are constructed out the diagonal generators $K_{R/L}^{a,m}$:
\begin{equation}
K_{R/L}^{a,m}=J_{R/L}^{a,m}+J_{R/L}^{\prime a,m}.
\end{equation}
The interactions that gap the diagonal subgroup are
\begin{equation}
\mathcal{L}_{su(2)}^{inter}=
-\sum_{m=1}^{N}\sum_{a=1}^3 g_ m^a K_{L}^{a,m} K_{R}^{a,m},
\end{equation}
since the coupling constants $g_m$ also flow to strong coupling at low energies \cite{Huang}.


\end{document}